\documentclass[aps,prd,preprintnumbers,showpacs]{revtex4}
\usepackage[dvips]{graphicx}
\begin{document}

%
%

\eprint{Nisho-1-2018}
\title{A phenomenological model of QCD monopole hadron interactions}
\author{Aiichi Iwazaki}
\affiliation{Nishogakusha University,\\ 
6-16 Sanbancho Chiyoda-ku Tokyo 102-8336, Japan.}   
\date{Sept. 1, 2018}
\begin{abstract}
Monopoles have recently been discussed to be a dominant component in strong coupled quark gluon plasma ( QGP )
and to play a role for chiral symmetry breaking as well as quark confinement.
We analyze monopole
quark interactions and show that  
massless quarks colliding with the monopoles inevitably change their chiralities keeping their flavors.
The monopole quark interaction explicitly breaks the chiral symmetry ( SU$_A(2)\times $U$_A$(1) ) just like bare quark masses.
It is given by $\bar{q}q\Phi^{\dagger}\Phi$ with the monopole field $\Phi$.
The pions are not Nambu-Goldstone bosons even in the vanishing bare quark masses. 
Their masses are mainly determined by the interaction because
the monopole condensation generates a larger current quark mass than the bare quark masses.
Based on the analysis of the monopole quark interaction,
we propose a phenomenological linear sigma model coupled with the monopoles.
The monopoles couple only with isoscalar scalar mesons e.g. sigma meson $\sigma$ 
such as $\sigma\Phi^{\dagger}\Phi$ 
indicated by the monopole quark interaction.
The coupling explicitly breaks the chiral SU(2)$_A\times $U$_A$(1) symmetry.
Pion masses are generated by the chiral condensate, which 
arises only when the monopole condensate takes place.
We show that one of the monopoles is a color singlet and observable.
The monopole decays into hadrons ( pions, kaon, etc. ) through the coupling.
Our analysis indicates that $f_0(1500)$ meson is a candidate of the observable monopole.    
As phenomenological effects of these monopoles, we point out that the masses of hadrons decrease in dense nuclear matters
and that  chiral magnetic effects disappear
in strong coupled QGP.  
\end{abstract}
\hspace*{0.3cm}

\hspace*{1cm}

\maketitle

\section{Introduction}

It is well known that monopoles in QCD play important roles for quark confinement. 
The monopole condensation produces dual superconductor in which color electric fields
are squeezed into vortices. That is, color electric flux tube is formed between a quark and an anti-quark.
The confinement picture is intuitively understandable so that  
the mechanism is favorable. It
has been extensively analyzed in lattice gauge theories. Especially, by using Maximal Abelian gauge,
the picture\cite{dual} of the confinement has been examined.
( There is another picture of the confinement based on Polyakov loop. It has been shown\cite{iwazaki} that the monopole condensation
leads to the vanishing Polyakov loop, which implies the quark confinement. ) 

The monopoles have recently been discussed\cite{hasegawa} to play another role for chiral symmetry breaking.
In particular, their association with zero modes of Dirac operators has been examined:
Addition of a pair of monopole and anti-monopole to gauge field configuration in Dirac operators
increases the number of the zero modes. 
It suggests that the monopoles play a role for the chiral symmetry breaking.
Recently, their observable effects have been discussed\cite{heavy} in the formation of quark gluon plasma ( QGP ).
The monopoles are quasi-particles in strong coupled QGP
and are discussed to be a dominant component of the plasma.  
Furthermore, it has been discussed that 
they play a role of the decay of glasma produced just after high energy heavy ion collisions.
In this way the roles of the monopoles in strong coupled QCD have been increasingly
recognized phenomenologically as well as theoretically.

Obviously they are glueballs, i.e.
composites of gluons. There would be many types of glueballs. The monopoles
are ones of such glueballs. Until now, there are no phenomenological models of monopole hadron interactions.
One of the reasons
is that we have no explicit ideas how they interact with hadrons. We have phenomenological models of hadrons such as 
sigma models or Nambu-Jona-Lasinio model, which are constructed based on chiral symmetry. Because we have no 
guiding principle for the construction of the monopole hadron interactions,  we do not have
phenomenological models of the monopoles ( glueballs ) interacting with hadrons.

\vspace{0.1cm}
In this paper, we present a phenomenological model of the monopoles coupled with hadrons.
It is constructed based on the analysis of the monopole quark interactions. 
The interaction is shown to explicitly break  the chiral symmetry ( SU$_A(2)\times $U$_A$(1) )
even if the bare quark masses vanish.

It is well known that we need to impose a boundary condition for the fermions at the location of the monopole
in the monopole fermion scattering.
The boundary condition violates the chiral symmetry. In other words, the massless fermions must flip their chiralities when
they are scattered by the monopole.
We may describe such a monopole fermion ( quark ) local interaction by the effective Lagrangian
such as $\bar{q}q\, \Phi^{\dagger}\Phi$ ( $q=u, d$ quarks ) using the monopole field $\Phi$. The chiral non symmetric interaction is
similar to the bare mass terms of the quarks.
Since the chiral symmetry ( SU$_A(2)\times $U$_A$(1) ) is explicitly broken in the interaction, pions are not Nambu-Goldstone bosons.
Their masses receive the contributions from the interaction.
We discuss that the interaction is weak in comparison with strong interactions between quarks and gluons.
The weakness of the interaction leads to small pion masses. 

\vspace{0.1cm}
Based on the analysis, we propose a phenomenological linear sigma model
coupled with the QCD monopoles. 
The monopoles interact only with isoscalar scalar mesons composed of $\bar{q}q$, e.g. sigma meson $\sigma$. 
For instance, the monopole hadron interaction is given by $\sigma\Phi^{\dagger}\Phi$.
( It corresponds to the monopole quark interaction $\bar{q}q\Phi^{\dagger}\Phi$. )
The interaction explicitly breaks the chiral symmetry and leads to the mixing between the monopoles and the isoscalar scalar mesons.
We show that the mixing is very weak.
The monopoles couple with other hadrons only through the mixing.
Thus, their decay widths are small.

The monopoles themselves are assumed to be described in
a model of dual superconductor\cite{dual,suzuki,che,suga}.
In the model the monopole condensation leads to the quark confinement.
On the other hand,
the chiral condensate arises in our model only when the monopole condensation takes place.
Without the monopole hadron interaction, our linear sigma model is chiral symmetric and does not show spontaneous chiral symmetry breaking.
Our sigma model does not have standard Wine bottle potential. 
The chiral condensate is generated only when the monopole condensation takes place.
We discuss some phenomenological implications expected from the model, e.g. $f_0(1520)$ meson is a candidate for the monopole. 

Our model is a hybrid model of a sigma model of hadrons and a dual superconducting model of monopoles.
It is an original one with the monopole hadron interaction, as long as we know.
We analyze the model only in a tree level. But the treatment may be fair 
as a first step of the analysis of a nonstandard model involving the monopole hadron
interaction.

\vspace{0.1cm}
In our previous paper\cite{iwazakinew} we have shown such a hybrid model, but the model
is assumed to be chiral symmetric. Thus, the pion is a Goldstone boson of 
the chiral symmetry which is broken spontaneously in the model.
The idea is very favorable.
But detailed analysis of monopole quark interactions shows that the presence of the monopoles explicitly breaks the chiral SU$_A$(2)
symmetry. Our point of view is that the chiral symmetry in strongly coupled QCD is explicitly broken in the presence of the monopoles
even if the bare quark masses vanish.

\vspace{0.1cm}

In order to treat the monopoles explicitly, we adopt the assumption of the Abelian dominance.
It only holds in the low energy phenomena of QCD, that is, strong coupled QCD.
The validity of the assumption has been discussed in the lattice gauge theory by using Maximal Abelian gauge.
According to the Abelian dominance, the relevant dynamical degrees of freedom are massless diagonal gluons, i.e. $A_{\mu}^{3,8}$, massless quarks
and the three types of the monopoles. The off diagonal gluons are massive and irrelevant to the low energy phenomena. 
An effective model describing quark confinement by the monopoles, is Dual superconducting model.
We adopt the effective model for simplicity.  
We point out that one of the monopoles is color singlet and observable in a phase of monopole condensation in which color magnetic charges
are screened. We find that
the monopole decays into pion, kaon, $\eta$, etc. through the isoscalar scalar mesons coupling e.g. $\sigma\Phi^{\dagger}\Phi$.

We predict some of phenomenological effects obtained by the analysis of the monopole quark interactions.
Because the chiralities of quarks are not conserved in the presence of the monopoles,
chiral magnetic effects may not arise in high energy heavy ion collisions. The effects are caused by
the chiral imbalance in QGP produced in the early stage of the collisions, i.e. the stage of glasma decay.
The chiral imbalance is washed out of QGP owing to the monopole excitations appearing near 
deconfinement transition temperature.
Furthermore, we predict that 
partial chiral restoration, i.e. the decrease of hadron masses may arise in dense nuclear matters.
This is because the monopole condensate
diminishes in the matters. Stronger color electric fields in the matters 
expel the monopole condensates than those do in hadrons. The chiral condensates diminish owing to the 
decrease of the monopole condensates. The chiral condensates are generated by the monopole condensation
in our model.

In the next section (\ref{boundary}), we discuss the monopole quark interactions and show that we need to impose a boundary condition
for quarks at the monopoles. The condition explicitly violates the chiral SU$_A(2)\times $U$_A$(1) symmetry. 
Thus, the pions are not Nambu-Goldstone bosons.
We show that the monopole
quark interaction is weak relative to the other hadronic interactions.
The weakness of the interaction causes the small pion masses. We discuss some phenomenological consequences
of the monopole quark interactions. 
Then, we propose a SU(2) linear sigma model coupled with the monopoles in section (\ref{su(2)}). The monopole hadron coupling 
explicitly breaks the chiral symmetry and mixes the monopoles with
isoscalar scalar mesons, i.e. sigma meson in the SU(2) model.
In section (\ref{monopole}) we identify an observable monopole as an invariant state under Weyl symmetry.
In section (\ref{small mixing}) we show that the monopole hadron interaction is small in the SU(2) model.
We also explicitly show the masses of the hadrons and the monopoles, and their mixing in the model.
In section (\ref{decay}) we discuss how the 
observable monopole decays into hadrons. The decay proceeds through a small component of the monopole in the sigma field,
which couples with pions in SU$_V$(2)$\times$SU$_A$(2) invariant way. 
In section ({\ref{su(3)}) we generalize a hybrid SU(2) linear sigma model to a hybrid SU(3) linear sigma model.
Summary and conclusion follow in the final section (\ref{con}).

\section{Chirality non conserved monopole quark interaction}
\label{boundary} 
First, we would like to show that the chiralities are not conserved in the monopole quark scattering. That is 
the chiral U$_A$(1) symmetry is broken. 

When we consider monopole quark interaction in QCD, we notice that the monopoles arise only in the low energy
region of QCD. That is, there are no color magnetic monopoles as dynamical degrees of freedom
in high energy region where perturbative QCD is valid. In the low energy region it is unclear what are
effective dynamical degrees of freedom relevant to understanding physics of QCD. The coupling strength is so large
as for quarks and gluons themselves to be not relevant dynamical degrees of freedom.
Generally, in the low energy region the assumption of Abelian dominance\cite{abelian,maximal} may hold.
According to the assumption, the dynamical degrees of freedom relevant to the 
low energy physics are massless quarks, maximal Abelian ( diagonal ) gauge fields and
monopoles. Off diagonal components of the gauge fields are massive so that they are not
relevant. Although the monopoles are massive, they are nearly massless at a critical temperature where they begin to condense in vacuum.
After their condensation, the masses of excited monopoles are much less than those of off diagonal gluons. 
Thus, relevant excitations in the low energy QCD are massless quarks, Abelian ( diagonal ) gluons and the monopoles.
In the present paper we assume the Abelian dominance in order to treat explicitly the monopole excitations in strong coupled QCD.

\vspace{0.1cm}
 Under the assumption we discuss monopole quark scattering.
In order to do so, we briefly explain monopole excitations in SU(3) gauge theory. 
In SU(3) gauge theory, we have three types of monopoles, which are characterized by root vectors of SU(3),
 $\vec{\epsilon}_1=(1,0), \vec{\epsilon}_2=(-1/2,-\sqrt{3}/2)$ and $\vec{\epsilon}_3=(-1/2,\sqrt{3}/2)$.
They describe the couplings with the maximal Abelian gauge fields, $A_{\mu}^{3,8}$ such as $\epsilon_i^a A_{\mu}^a$.
For example a monopole with $\vec{\epsilon}_1$ couples only with $A_{\mu}^3=\epsilon_1^a A_{\mu}^a$. Thus, the quarks coupled with the monopole
are a doublet $q=(q^+,q^-,0)$ of the color triplet. Here the index $\pm$ of $q^{\pm}$ denotes a positive or negative charged component associated
with the gauge field $A_{\mu}^3\lambda^3/2$; $\lambda^a$ are Gell-Man matrices.
Similarly the other monopoles couple with the quark doublets, $q=(q^+,0,q^-)$ and $q=(0,q^+,q^-)$. 
Thus, we consider a massless quark doublet $\Big(\begin{array}{l}q^+ \\ q^-\end{array} \Big)$ scattered by the monopole, 

\begin{equation}
\label{1}
\gamma_{\mu}(i\partial^{\mu}\mp \frac{g}{2}A^{\mu})q^{\pm}=0,
\end{equation}
where the gauge potentials $A^{\mu}$ denotes a Dirac monopole
given by 

\begin{equation}
\label{2}
A_{\phi}=g_m(1-\cos(\theta)), \quad A_0=A_r=A_{\theta}=0
\end{equation} 
where $\vec{A}\cdot d\vec{x}=A_rdr+A_{\theta}d\theta+A_{\phi}d\phi$ with polar coordinates $r,\theta$ and $\phi=\arctan(y/x)$.
$g_m$ denotes a magnetic charge with which magnetic field is given by $\vec{B}=g_m\vec{r}/r^3$.
The magnetic charge satisfies the Dirac quantization condition $g_mg=n/2$ with integer $n$ where $g$ denotes the gauge coupling of SU(3) gauge theory.
Hereafter, we assume the monopoles with the magnetic charge $g_m=1/2g$.

\vspace{0.1cm}

The monopole quark ( in general, fermion ) dynamics has been extensively explored, in particular, in the 
monopole catalysis\cite{rubakov,callan,ezawa} of baryon decay ( so called Rubakov effect ).
The point is that conserved angular momentum has an additional component. That is,
it is given by

\begin{equation}
\label{3}
\vec{J}=\vec{L}+\vec{S}\mp gg_m\vec{r}/r
\end{equation}
where $\vec{L}$ ( $\vec{S}$ ) denotes orbital ( spin ) angular momentum of quark.
The additional term $\pm gg_m\vec{r}/r$ play an important role of chiral symmetry breaking.
Owing to the term we can show that either the charge or the chirality is not conserved in the monopole quark scattering.
When the chirality ( or helicity $\sim \vec{p}\cdot\vec{S}/|\vec{p}||\vec{S}|$ ) is conserved, the spin must flip $\vec{S}\to -\vec{S}$ after the scattering 
because the momentum flips after the scattering; $\vec{p}\to -\vec{p}$.
Then, the charge must flip $g\to -g$ because of the conservation of $\vec{J}\cdot\vec{r}$, i.e.
 $\Delta(\vec{J}\cdot\vec{r})=\Delta(\vec{S}\cdot\vec{r})+\Delta(gg_mr)=0$. ( $\Delta(Q)$ denotes the change of the value $Q$ after the scattering. ) 
On the other hand, when the charge is conserved ( it leads to $0=\Delta(\vec{J}\cdot\vec{r})=\Delta(\vec{S}\cdot\vec{r}$) ),
the chirality $\vec{p}\cdot\vec{S}/|\vec{p}||\vec{S}|$ must flip because the spin does not flip $\vec{S}\to \vec{S}$.
Thus we find that either the charge or the chirality conservation is lost in the scattering.
( In the discussion we assume that the mass of the monopole is sufficiently large such that the collision of the quark
does not change the state of the monopole. That is, the energy of the quark is much less than the mass of the monopole. )

\vspace{0.1cm}
The charge is strictly conserved because the charge conservation is guaranteed by the gauge symmetry.
When the quark flips its charge, monopoles are charged or heavy charged off diagonal gluons must be produced to preserve the charge conservation.
But the processes cannot arise in the low energy scattering. Charged monopoles are dyons and they are heavy.   	
Thus, inevitably the chirality is not conserved. The right handed quark $q_R$ is transformed to the left handed quark $q_L$ in the scattering.
That is, 
we impose a boundary conditions $q^{\pm}_R(r=0)=q^{\pm}_L(r=0)$ at the location of the monopole, which breaks the chiral U$_A$(1) symmetry.
The chiral symmetry breaking has been rigorously shown\cite{rubakov,ezawa} to be caused by chiral anomaly in QCD.
( Even if we impose chirality conserved but charge non conserved boundary conditions $q^+_{R,L}(r=0)=q^-_{R,L}(r=0)$ at the location of the monopole, 
we can show\cite{rubakov,ezawa} that the charge is conserved,
but chirality is not conserved in the monopole quark scattering. 
The chirality non conservation arises from chiral condensate $\langle\bar{q}q\rangle\propto 1/r^3$
locally present around each monopole at $r=0$.
The condensate is formed by the chiral anomaly when we take into account 
quantum effects of gauge fields $A_{\mu}=\delta A_{\mu}^{quantum}+A_{\mu}^{monopole}$.  
Eventually, the chirality non conserved boundary condition is realized in physical processes. 
These results are by-products in the analysis of the Rubakov effect\cite{rubakov, ezawa}. )

Furthermore, it apparently seems that
quarks may change their flavors in the scattering. For instance, u quark is transformed into d quark.
Then, the monopole must have a SU(2) flavor after the scattering. But it is impossible because there are no such monopoles
with SU(2) flavors in QCD. They are flavor singlet.
Therefore, quarks cannot change their flavors in the scattering with the monopole.
Quarks change only their chiralities. It results in SU$_A$(2) chiral symmetry breaking in the scattering.
The symmetry is defined such that flavor SU(2) quark doublet $q=\Big(\begin{array}{l}u \\  d\end{array} \Big)$
is transformed into $\exp(i\gamma_5\vec{\theta}\cdot\vec{\tau})q$ in which
$\vec{\tau}$ denotes generators of the flavor SU(2). If the chiral SU$_A$(2) symmetry is exact in the scattering, the flavor of quark may change when
the quark changes its chirality. But this does not take place. Thus, the chiral SU$_A$(2) symmetry is broken.

\vspace{0.1cm}
The chiral symmetry breaking arises at the location of the monopole. It is associated with properties at short distances.
So it is beyond the assumption of Abelian dominance.
But, the explicit symmetry breaking is not artifact of the assumption.
In QCD we have monopole solutions such as Wu-Yang monopoles. When we admit the presence of the monopoles
in QCD, we need appropriate boundary conditions\cite{kazama} for quarks at the locations of the monopoles
in order to quantum mechanically define the monopole quark scattering. The relevant boundary conditions are those such as boundary conditions 
breaking the chiral SU$_A$(2) symmetry as well as U$_A$(1) symmetry. 

\vspace{0.1cm}
In high energy region, perturbative analysis of QCD works well so that
the chiral symmetry is preserved.
But it is not preserved in low energy regions where the monopole excitations explicitly break the chiral symmetry. 
In other wards the chiral symmetry holds in high temperature, 
while it is explicitly broken in low temperature even in the vanishing bare quark masses. 
The chiral symmetry is not spontaneously broken but explicitly broken in the low temperature. 

\vspace{0.1cm}
It apparently seems that this is different from standard understanding; the chiral symmetry is spontaneously broken
in low temperatures.
But we should note that the bare mass terms of light quarks explicitly break the chiral symmetry.
Because the bare masses are small compared with standard hadronic scales $\sim 100$MeV, 
the chiral symmetry approximately holds and it is spontaneously broken.
The present case is similar to the case in the bare mass terms.
The monopole quark interaction is weak compared with hadronic interactions as we show just below.
Thus, the chiral symmetry approximately holds.

It is easy to see the smallness of the monopole quark interaction.
The interaction is not controlled by the strong coupling $\alpha_s=g^2/4\pi$ of QCD.
This is because the quarks have color charges $g$ and the monopoles have magnetic charges $g_m\propto 1/g$.
Therefore, the monopole quark interaction ( $\propto g\times g_m$ ) 
does not involve $\alpha_s$ at tree level. On the other hands, the quark gluon interactions involve
$\alpha_s$ even at tree level.  
It implies that the monopole quark interaction is weak compared with the other hadronic interactions. 
It would be of the order of $gg_m/g^2=1/8\pi\alpha_s$.  
Therefore,  
pion masses receiving the effects are small because the interaction is weak. 

But a distinction is present.
When the bare quark mass terms and the monopole quark interaction terms are absent, the chiral symmetry is exact.
According to the standard idea, the chiral symmetry is spontaneously broken and
the chiral condensate arises. However,
the mechanism of the spontaneous symmetry breaking or the generation mechanism of the chiral condensate is unclear. 
On the other hand, according to our analysis,
the chiral symmetry is not spontaneously broken even in the absence of
the monopole quark interaction. The chiral symmetry is not broken spontaneously.
Our idea is that
the monopole condensation generates
the chiral condensate $\langle\bar{q}q\rangle\neq 0$. 
This is because each monopoles carries local chiral condensate $\langle\bar{q}q\rangle\propto 1/r^3$ \cite{ezawa}.
Therefore, once the monopole quark interaction is switched on, the chiral symmetry is explicitly broken and
the chiral condensate arises simultaneously when the monopoles condense.
Actually,
we have recently shown\cite{iwazaki2} by using the effective monopole quark interaction $\bar{q}q\Phi^{\dagger}\Phi$ that
the chiral condensate $\langle\bar{q}q\rangle\neq 0$ arises only when the monopoles condense $\langle\Phi\rangle\neq 0$.
Our idea is consistent with the recent works\cite{hasegawa}, which show that the presence of the monopoles
induces zero modes of Dirac operators.  The increase of the number of the zero modes suggests the presence of
the chiral condensate.  
The generation mechanism of the chiral condensate is obvious in our case, while
it is unclear in the standard analyses of the chiral symmetry breaking.


\vspace{0.1cm}

Quarks change their chirality without the change of their flavors in the monopole quark scattering;
the chiralities change at the location of the monopole.
Effectively, we can describe such a interaction by using monopole field $\Phi$ that
$-g'|\Phi|^2(\bar{u}_Lu_R+\bar{u}_Ru_L+\bar{d}_Ld_R+\bar{d}_Rd_L)=-g'|\Phi|^2(\bar{u}u+\bar{d}d)$ with $g'>0$. 
( More precisely, the monopole quark interaction is described by using three types of the monopole fields $\Phi_i$ ($i=1\sim 3$ ) such that 
$-g'(|\Phi_1|^2(\bar{q}_1q_1+\bar{q}_2q_2)+|\Phi_2|^2(\bar{q}_1q_1+\bar{q}_3q_3)+|\Phi_3|^2(\bar{q}_2q_2+\bar{q}_3q_3))$
where indices $i$ of  $q_i$ denotes color component of quark $q=u$ or $d$. The interaction preserves Weyl symmetry discussed below. )
The parameter $g'$ is of the order of the inverse of $\Lambda_{QCD}$ times $gg_m/g^2=1/8\pi\alpha_s$; 
$g'\simeq (1/8\pi\alpha_s)\Lambda_{QCD}^{-1}$. Numerically, it is given such that $g'\sim (3100\mbox{MeV})^{-1}$
for $\Lambda_{QCD}=250$MeV and $\alpha_s(Q=1\mbox{GeV})\simeq 0.5$.

Obviously the interaction explicitly breaks 
chiral SU$_A$(2) symmetry. 
Thus, pions are not Nambu-Goldstone bosons. Their masses are generated 
by the chiral non symmetric monopole quark interaction and the bare quark masses.
We also note that the interaction mixes the monopoles with isoscalar mesons composed of $\bar{q}q$.
The monopoles are glueballs so that the coupling gives the explicit way of the mixing among the glueballs and isoscalar mesons.  

It should be noted that the interaction term explicitly contributes to quark masses $m_q$ when the monopoles condense $\langle\Phi\rangle\neq 0$.
( More precisely, $-2g'|\Phi_0|^2(\bar{q}_1q_1+\bar{q}_2q_2+\bar{q}_3q_3)=-2g'|\Phi_0|^2\bar{q}q$ with $\Phi_0\equiv\langle\Phi_i\rangle$. 
So, $m_q=2g'|\Phi_0|^2$. )
Thus, the current quark masses are 
given by the bare quark masses and the quark masses generated by the monopole condensation.
Gell-Mann-Oaks-Renner relation is modified such as $m_{\pi}^2f_{\pi}^2=-(2m_{ud}+3g'\Phi_0^2)\langle\bar{q}q\rangle$. 
The relation indicates that the pion masses are mainly generated by the monopole condensation, not their bare masses.
Indeed, the coefficient $3g'\Phi_0^2/2$ is much larger than the bare quark masses $m_{ud}\sim 5$MeV. For instance,
$3g'\Phi_0^2/2\sim 50$MeV ( $20$MeV ) for $\Phi_0\sim \Lambda_{QCD}=250$MeV ( $\Phi_0\sim 170$MeV, see later).

Furthermore, the monopole condensate is reduced in dense nuclear matter because the condensate is expelled by color electric
fields; fluctuations of color electric fields are larger in the dense nuclear matter than those in hadrons. 
Then, we expect that the masses of the hadrons in the dense nuclear matter are smaller than those in vacuum.


\vspace{0.1cm}
The chiralities of quarks are not conserved when they scatter with the monopoles. The fact gives rise to an important result 
in strong coupled QGP produced by high energy heavy ion collisions. It has recently been pointed out that chiral magnetic effects\cite{chiralmagnetic} can be seen in the plasma.
The effects are caused by chiral imbalance  (the number of quarks with positive chirality
minus the number of quarks with negative chirality )
in the plasma. The imbalance is produced in the early stage of the high energy heavy ion collisions. It is the stage of glasma decay. 
The imbalance is represented by chiral chemical potential $\mu_5$. 
The nonzero chiral chemical potential $\mu_5\neq 0$ leads to electric currents parallel to magnetic fields produced in the ionized heavy ion collisions.
The magnitudes of the currents are proportional to $\mu_5$.
It is called as chiral magnetic effect. In QGP with temperatures near the crossover at which quark-hadron phase transition takes place,
the monopoles have been shown\cite{heavy} to be dominant components of the plasma.  
Then, the chiral imbalance $\mu_5\neq 0$ is washed out by the monopoles.
That is, it is realized that $\mu_5=0$.
Therefore, the chiral magnetic effects cannot be observed; the chiral imbalance vanishes in the monopole dominant phase of the QGP.

\section{Linear sigma model coupled with monopole}
\label{su(2)}
Based on the analysis of the monopole quark interaction,
we would like to propose a hybrid model of a SU(2) linear sigma model and a model\cite{maedan,suga,che} of dual superconductor.

The both models are described by 

\begin{eqnarray}
\label{4}
L_{\Sigma}&=&\frac{1}{4}\rm{Tr}(\partial_{\mu}\Sigma^{\dagger}\partial^{\mu}\Sigma) 
-\frac{m^2}{4}\rm{Tr}(\Sigma^{\dagger}\Sigma)
-\frac{\lambda}{4} (\rm{Tr}(\Sigma^{\dagger}\Sigma))^2
=\frac{1}{2}(\partial\sigma)^2+\frac{1}{2}(\partial\vec{\pi})^2 \nonumber \\
&-&\frac{m^2}{2}(\sigma^2+\vec{\pi}^2)-\lambda(\sigma^2+\vec{\pi}^2)^2,
\end{eqnarray}
with  $\Sigma\equiv \sigma +i\vec{\pi}\cdot\vec{\tau}$ and $m^2, \lambda>0$,

and 

\begin{equation}
\label{5}
L_{\Phi}=\sum_{i=1\sim 3}(\frac{1}{2} |D_{\mu}^i\Phi_i|^2+\mu^2|\Phi_i|^2-\lambda' |\Phi_i|^4)-\frac{1}{4}((B^3_{\mu\nu})^2+(B^8_{\mu\nu})^2)
\end{equation}
with $B^a_{\mu\nu}=\partial_{\mu}B^a_{\nu}-\partial_{\nu}B^a_{\mu}$ and $\mu^2,\lambda'>0$,
where we have denoted
three types of the monopoles $\Phi_i$ and two types of dual gauge fields $B_{\mu}^a$ ( $a=3,8$ );
$D_{\mu}^i\equiv \partial_{\mu}-ig'_m\epsilon_i^a B_{\mu}^a$ 
with root vectors $\vec{\epsilon}_i$ of SU(3) and $g'_m=4\pi/g$.
There are dual gauge symmetry such that $B^a_{\mu}\to B^a_{\mu}+\partial_{\mu}\Lambda^a$ and $\Phi_i\to \Phi_i\exp(ig'_m\epsilon^a_i\Lambda^a)$.
The gauge symmetry is dual one to the Maximal Abelian gauge symmetry of SU(3), which is the remaining gauge symmetry after taking
a Maximal Abelian gauge\cite{maximal} in color SU(3) gauge theory. 

\vspace{0.1cm}
The linear sigma model is a chiral SU$_V$(2)$\times$ SU$_A$(2) symmetric model of hadrons ( pion and sigma ).
The parameter $m^2$ is taken to be negative in the standard sigma model. But in our model it is taken to be positive.
Thus, spontaneous chiral symmetry breaking does not take place. 
There is no condensation of the sigma field $\sigma$
without monopole sigma interaction. When we introduce
an interaction between the monopoles $\Phi_i$ and sigma meson $\sigma$, 
the condensation $\langle \sigma \rangle\neq 0 $ of the sigma field $\sigma$ takes place. In particular,
it does only when the monopole condensation $\langle \Phi \rangle \neq 0$ arises.
That is, the confinement ( monopole condensation ) and the chiral symmetry breaking ( chiral condensation )
simultaneously occur.  Here we make a comment that it is still unclear why the monopoles condense, although
we clarify why the chiral condensate appears.

As explained above, when quarks collide with monopoles, their chiralities are not conserved
owing to the chiral non conserved boundary condition at the location of the monopole.
The collisions is
effectively described by a Lagrangian $-g'|\Phi|^2\bar{q}q$.
Thus, it is natural to take an interaction between the monopoles $\Phi_i$ and the sigma meson $\sigma$ in the following,

\begin{equation} 
\label{6}
L_{\rm int}=h\sigma \sum_{i=1,2,3} |\Phi_i|^2
\end{equation}
where the dimensional parameter $h$ describes strength of the coupling. 

Therefore, our hybrid model of
the linear sigma model and the model of dual superconductor is described by the Lagrangian, $L_{\Sigma}+L_{\Phi}+L_{\rm int}$.
Obviously the interaction $L_{\rm int}$ explicitly breaks the chiral SU$_A$(2) symmetry.
Thus, there are no Nambu Goldstone bosons associated with the symmetry.
Pions acquire their masses receiving the effects of this interaction term. We expect that the term is small relative to the other interaction terms.
The smallness of the mixing term comes from the weak monopole quark interaction mentioned above. 


The mixing among the monopoles $\Phi_i$ and the sigma meson $\sigma$ are described by the interaction $L_{\rm int}$.
Furthermore, the monopoles decay into ordinary hadrons only through this mixing term.
When the mixing term is small, their decay widths would be small.
( When $h=0$, the monopoles do not decay into hadrons. )
Namely,
the sigma field $\sigma$ involves small components of the physical monopole and couples with ordinary hadrons as described by the Lagrangian $L_{\Sigma}$.
Thus, the decay width of the monopole is small when the mixing parameter $h$ is small.

Intuitively,
the monopoles could decay into gluons which subsequently form ordinary hadrons.
That is, the monopoles decay into the ordinary hadrons. But the decay is forbidden when Abelian dominance is valid.
Because the gluons produced by the monopole decay are massive off diagonal gluons, the decay is suppressed.
Actually, the dual superconducting model does not allow the monopole decay. The monopoles are stable.
The monopoles can decay only when they couple with quarks by the interactions $g'|\Phi|^2\bar{q}q$.


\vspace{0.1cm}
As we show below, the mixing parameter $h$ is small; it is of the order of $10$MeV
much smaller than the typical hadronic scale $100$MeV. 
Similar small mixing of the glueball and the isoscalar scalar mesons has been shown\cite{glue} in lattice gauge theory.

\section{Physical states of monopoles}
\label{monopole}
In order to examine the observational effects of monopoles, we need to find observable monopoles
which should be color singlet. There are three types of monopoles, $\Phi_i$ ( $i=1,2,3$ ) in the model of color SU(3) dual superconductor.
The model is obtained under the assumption of Abelian dominance, by using Maximal Abelian gauge.
The Abelian gauge symmetries still remain. Physical states should be neutral associated with the gauge symmetries.
Additionally,
a discrete symmetry of color SU(3) 
still remains in the model. It is a Weyl symmetry.
The physical observable should be invariant under the discrete symmetry.
The symmetry is defined such that
the three types of the colors in quarks $q_1=(1,0,0)$, $q_2=(0,1,0)$ and $q_3=(0,0,1)$ are
transformed into each other under the elements $U_a$ of SU(3), $U_{(1,2)}q_1=q_2$, $U_{(1,3)}q_1=q_3$, and $U_{(2,3)}q_2=q_3$.
Using the elements, $B^3_{\mu}\lambda_3+B^8_{\mu}\lambda_8$ is transformed into 
$U_a^{\dagger}(B^3_{\mu}\lambda_3+B^8_{\mu}\lambda_8)U_a=B'^3_{\mu}\lambda_3+B'^8_{\mu}\lambda_8$;
$\lambda_3$ and $\lambda_8$ denote diagonal Gell-Man matrices. 
That is, we obtain the transformation, $B^3_{\mu}\to B'^3_{\mu}$ and $B^8_{\mu}\to B'^8_{\mu}$,

\begin{eqnarray}
\label{7}
(&B^3_{\mu}& \to -B^3_{\mu}, \,\,\,B^8_{\mu}\to B^8_{\mu} \quad),\quad 
(\quad B^3_{\mu} \to \frac{1}{2}B^3_{\mu}-\frac{\sqrt{3}}{2}B^8_{\mu}, \,\,\,B^8_{\mu}\to
-\frac{\sqrt{3}}{2}B^3_{\mu}-\frac{1}{2}B^8_{\mu}\quad) \nonumber \\
(&B^3_{\mu}& \to \frac{1}{2}B^3_{\mu}+\frac{\sqrt{3}}{2}B^8_{\mu},\,\,\,
B^8_{\mu}\to \frac{\sqrt{3}}{2}B^3_{\mu}-\frac{1}{2}B^8_{\mu}\quad).
\end{eqnarray}

Because the dual gauge fields $B^a_{\mu}$ minimally couple with the monopoles $\Phi_i$ 
such as $D^i_{\mu}\Phi_i=(\partial_{\mu}-ig_m\vec{\epsilon}_i\cdot\vec{B}_{\mu})\Phi_i $,
the Weyl invariance requires that the monopole fields are transformed under the transformation in eq(\ref{7}) in the following,

\begin{eqnarray}
&(&\,\, \Phi_1\to \Phi^{\dagger}_1,\,\,\,\Phi_2\to \Phi^{\dagger}_3,\,\,\, \Phi_3\to \Phi^{\dagger}_2\,\, ),\,\,
(\,\, \Phi_1\to \Phi^{\dagger}_3,\,\,\,\Phi_2\to \Phi^{\dagger}_2,\,\,\, \Phi_3\to \Phi^{\dagger}_1\,\, ), \nonumber \\ 
&(&\,\, \Phi_1\to \Phi^{\dagger}_2,\,\,\,\Phi_2\to \Phi^{\dagger}_1,\,\,\, \Phi_3\to \Phi^{\dagger}_3\,\, )
\end{eqnarray}
respectively. We find that the magnetic charge $\rho_m=\sum_{j=1,2,3}\Phi^{\dagger}_j(i\partial_0-g_m\vec{\epsilon}_j\cdot \vec{B}_0)\Phi_j+c.c.$
changes its sign under the transformation in eq(\ref{5}) and eq(\ref{6}); $\rho_m\to -\rho_m$.
Thus, the magnetic charge is not observable.

According to the transformation in eq(\ref{7}),
the observable monopoles are composites with no magnetic charges such as $\sum_{i=1,2,3} |\Phi_i|^2$
in the phase with no monopole condensation $\langle \Phi_i \rangle=0$.
On the other hand, there is an observable state $\sum_{i=1,2,3}\delta\Phi_i$
in the phase with monopole condensation $\langle\Phi_i\rangle=v\neq 0$
because we can take the fields $\delta\Phi_i$ real with $\Phi_i=v+\delta\Phi_i$.  
Therefore, we find that the monopole excitation $\frac{\sum_{i=1,2,3}\delta\Phi_i}{\sqrt{3}}$ is
an observable glueball in the confinement phase. The magnetic charge is screened in the phase. 
The monopole decays into hadrons through the mixing term.
On the other hand, $\phi_1\equiv \frac{\delta\Phi_1+\delta\Phi_2-2\delta\Phi_3}{\sqrt{6}} $ and $\phi_2\equiv \frac{\delta\Phi_1-\delta\Phi_2}{\sqrt{2}}$ 
represent color non singlet components of the monopoles. They are not observable.

\section{Sigma monopole mixing}
\label{small mixing}
Now we would like to show that the mixing term $\propto h$ in eq(\ref{6}) is small.
We first examine field configurations of ground states by solving field equations derived from the Lagrangian $L_{\Sigma}+L_{\Phi}+L_{\rm int}$,

\begin{eqnarray}
\label{9}
m^2\sigma_0+4\lambda \sigma_0^3=3h\Phi_0^2 \nonumber \\
-\mu^2+2\lambda'\Phi_0^2=h\sigma_0 ,
\end{eqnarray}
where $\langle \Phi_i \rangle =\Phi_0$ and $\langle \sigma \rangle =\sigma_0$.
The pions $\vec{\pi}$ has no vacuum expectation values. We note positive mass term $m^2>0$. 
Then, it follows that sigma meson has no vacuum expectation
value as long as $\Phi_0=0$. 
On the other hand, the sigma field can have vacuum expectation value when $\Phi_0\neq 0$.
Therefore, the chiral condensate takes place simultaneously when 
the monopole condensation $\Phi_0\neq 0$ does. 
The fact gives rise to a very important result that the chiral condensate $\langle\sigma\rangle$ locally diminishes in a region where
the monopole condensation $\langle\Phi_i\rangle$ diminishes. 
For instance, $\langle\Phi_i\rangle$ diminishes in a color electric flux tube between a quark and anti-quark.
The color electric field expels the monopoles just as magnetic field expels Cooper pairs in superconductor.
Then, it follows that the chiral condensate $\langle\sigma\rangle$ also diminishes in the flux tube. 
Actually, it has been shown\cite{miyamura,hasegawa} in lattice gauge theory that the chiral condensate  $\langle\bar{q}q\rangle$ diminishes in such a flux tube.
It is expected that a partial restoration of chiral symmetry in dense nuclear matter arises; mesons may be lighter in the nuclear matter than in vacuum.
We intuitively understand it such that
the chiral condensate diminishes in the matter 
because the monopole condensate inside dense nuclear matter is smaller owing to stronger color electric fields than 
that in vacuum.  

The result is coincident with the analysis of the monopole quark interaction $g'|\Phi|^2\bar{q}q$. 
The monopole condensation generates the current quark masses 
so that the diminishing of the condensate in dense nuclear matter leads to the diminishing of meson masses.

\vspace{0.1cm}
We proceed to show the smallness of the mixing parameter $h$.
We derive the pion mass,

\begin{equation}
\label{10}
m_{\pi}^2=m^2+4\lambda\sigma_0^2=\frac{3h\Phi_0^2}{\sigma_0}
\end{equation}
where we used the first equation in eq(\ref{9}). Obviously, the pion is not massless because the chiral SU(2)$_A$ symmetry is explicitly broken by
the interaction $L_{\rm int}$. Using the formula we can estimate the parameter $h$. 
We use the value of the parameter $\Phi_0\simeq 170$MeV previously estimated in the model of dual superconductor\cite{suga,che}. 
( The value was derived by neglecting contributions from the sigma model, i.e. $h=0$. )
Then, by noting the pion decay constant $f_{\pi}=\sigma_0\simeq 92$MeV and $m_{\pi}=135$MeV, we obtain $h\sim 20$MeV.
Therefore, we find that the monopole sigma meson mixing is very small in comparison to the typical hadronic scale $\sim 100$MeV.
( It confirms the consistency of the use of the value $\Phi_0=170$MeV which is obtained by neglecting the mixing. )

The small mixing parameter obtained by using phenomenological values is consistent with
the weak monopole quark interaction. The weakness of the interaction is theoretically derived. That is,
the strong coupling $\alpha_s$ does not appear in the monopole quark scattering at tree level,
while it arises in the quark gluon scatterings at tree level.
Thus, we find that the interaction term in eq(\ref{6}) of the monopoles and hadrons is consistent with the basic
monopole quark interaction.

%
%
%
%

\vspace{0.1cm}
We proceed to derive the mass eigenstates of sigma meson and monopoles.
By shifting the fields $\sigma$ and $\Phi_i$ such that $\sigma=\sigma_0+\delta\sigma$ and $\Phi_i=\Phi_0+\delta\Phi_i$
in the Lagrangian, we take only quadratic terms of these fields and $\vec{\pi}$,

\begin{eqnarray}
L&=&(\frac{m^2}{2}+6\lambda\sigma_0^2)(\delta\sigma)^2+(6\Phi_0^2\lambda'-\mu^2-h\sigma_0)\sum_{i=1\sim 3}(\delta\Phi_i)^2 \nonumber \\
&-&2h\Phi_0\delta\sigma\sum_{i=1\sim 3}\delta\Phi_i+\frac{m^2+4\lambda\sigma_0^2}{2}\vec{\pi}^2 \nonumber \\
&=&\frac{M^2}{2}\sum_{i=1,2}\phi^2_i+\frac{m_{sigma}^2}{2}\phi_{sigma}^2+\frac{m^2_{mon}}{2}\phi^2_{mon}+\frac{m_{\pi}^2}{2}\vec{\pi}^2,
\end{eqnarray}
with the eigenvalues

\begin{eqnarray} 
M^2&=&4(\mu^2+h\sigma_0), \quad m_{mon}^2=\frac{m^2+12\lambda\sigma_0^2+M^2+\sqrt{(m^2+12\lambda \sigma_0^2-M^2)^2+48h^2\Phi_0^2}}{2} \nonumber \\
m_{sigma}^2&=&\frac{m^2+12\lambda\sigma_0^2+M^2-\sqrt{(m^2+12\lambda\sigma_0^2-M^2)^2+48h^2\Phi_0^2}}{2},
\end{eqnarray}
where the eigenstates are given in the following,

\begin{eqnarray}
\phi_1&=&\frac{\delta\Phi_1+\delta\Phi_2-2\delta\Phi_3}{\sqrt{6}}, \quad \phi_2=\frac{\delta\Phi_1-\delta\Phi_2}{\sqrt{2}}, 
\quad \phi_{mon}=-\delta\sigma\sin\theta+\frac{\sum_{i=1\sim3}\delta\Phi_i}{\sqrt{3}}\cos\theta \nonumber \\
\phi_{sigma}&=&\delta\sigma\cos\theta+\frac{\sum_{i=1\sim3}\delta\Phi_i}{\sqrt{3}}\sin\theta
\end{eqnarray}
with the mixing angle is 

\begin{eqnarray}
\sin^2\theta&=&\frac{m^2+12\lambda\sigma_0^2-M^2+\sqrt{(m^2+12\lambda\sigma_0^2-M^2)^2+48h^2\Phi_0^2}}{2\sqrt{(m^2+12\lambda\sigma_0^2-M^2)^2+48h^2\Phi_0^2}} 
\nonumber \\
&=&\frac{m^2+12\lambda\sigma_0^2-m_{sigma}^2}{m_{mon}^2-m_{sigma}^2}  \to 0 \quad \mbox{as} \quad h\to 0
\end{eqnarray}
where we have assumed $4\mu^2\ge m^2$.

These mass eigenstates $\phi_{mon}$ and $\phi_{sigma}$ are physical observable.
We find that the sigma field $\sigma$ involves the observable monopole $\phi_{mon}$ as well as
the sigma $\phi_{sigma}$.
Similarly, the color singlet monopole field $\delta\Phi_1+\delta\Phi_2+\delta\Phi_3$ involves both components
$\phi_{mon}$ and $\phi_{sigma}$, That is,

\begin{equation}
\delta\sigma=\phi_{sigma}\cos\theta-\phi_{mon}\sin\theta, \quad \frac{\delta\Phi_1+\delta\Phi_2+\delta\Phi_3}{\sqrt{3}}=\phi_{mon}\cos\theta+\phi_{sigma}\sin\theta.
\end{equation}

We find that the sigma $\sigma$ involves small component ( $\propto \sin\theta $ ) of the physical monopole $\phi_{mon}$.
It leads to the small decay width of the monopole. 
 
\vspace{0.1cm}
When we generalize the model to  
the flavor SU(3) linear sigma model, we find that isoscalar mesons $\sigma$ and $f$ are mixed with monopoles;
the monopoles couple with the mesons such as $L_{\rm int}=h(\sigma_x+\sqrt{2}\sigma_y) \sum_{i=1,2,3} |\Phi_i|^2$.  
Therefore, an observable monopole decays into pions, kaons, and $\eta$ mesons through the 
interaction of $\sigma_x$ and $\sigma_y$ with the hadrons. Main decay modes are $\pi,\pi$ or $4\pi$ while secondary one is $K\bar{K}$.
Hence,
it is natural to identify the monopole as isoscalar $f_0(1500)$ meson.  

\section{Monopole decay}
\label{decay}
We can easily see how the monopole decays into pions. By taking interaction terms composed both of $\sigma$ and $\pi$ in the Lagrangian,

we obtain

\begin{eqnarray}
\label{15}
-\lambda((\sigma_0+\delta\sigma)^2+\vec{\pi}^2)^2 &\to& L_{\pi,\sigma} \nonumber \\
L_{\pi,\sigma}&\equiv&-\lambda(4\sigma_0 \delta\sigma \vec{\pi}^2+2(\delta\sigma)^2\vec{\pi}^2
+4\sigma_0(\delta\sigma)^3+(\delta\sigma)^4+\vec{\pi}^4)
\end{eqnarray}
where $\delta\sigma$ field is composed of observable sigma $\phi_{sigma}$ and monopole $\phi_{mon}$;
$\delta\sigma=\phi_{sigma}\cos\theta-\phi_{mon}\sin\theta$.  
On the other hand, we obtain interaction terms from $L_{\rm int}=h\sigma\sum_{i=1\sim 3}|\Phi_i|^2$,

\begin{eqnarray}
\label{16}
h(\sigma_0+\delta\sigma)\sum_{i=1\sim 3}(\Phi_0+\delta\Phi_i)^2 &\to& L_{\sigma,\Phi} \nonumber \\
 L_{\sigma,\Phi}\equiv h\delta\sigma\sum_{i=1\sim 3}(\delta\Phi_i)^2&=&
h\delta\sigma((\frac{\delta\Phi_1+\delta\Phi_2+\delta\Phi_3}{\sqrt{3}})^2+\phi_1^2+\phi_2^2).
\end{eqnarray} 

The Lagrangian $L_{\pi,\sigma}$ in eq(\ref{15})  represents interactions between the pions $\vec{\pi}$ and sigma $\sigma$, while
the Lagrangian $L_{\sigma,\Phi}$ represents interactions between sigma $\sigma$ and monopoles $\delta\Phi_i$.

\vspace{0.1cm}
Now we wish to discuss amplitudes of two pion decay of the monopole $\phi_{mon}$.
We can see that the tree amplitude of the monopole decay 
is proportional to $4\lambda\sigma_0\sin\theta$ caused by the term $4\lambda\sigma_0 \delta\sigma \vec{\pi}^2$ in $L_{\pi,\sigma}$
and there is no tree amplitude of the decay in the interaction $L_{\sigma,\Phi}$.
On the other hand, two one loop amplitudes of the decay are proportional to $4\lambda\sigma_0\sin\theta(2\lambda\cos^4\theta)$ and  
$4\lambda\sigma_0\sin^3\theta(2\lambda\sin^2\theta)$, respectively 
caused by the terms $4\lambda\sigma_0(\delta\sigma)^3$ and $2\lambda(\delta\sigma)^2\vec{\pi}^2$ in $L_{\pi,\sigma}$.
Furthermore, 
combining the terms $h\delta\sigma(\delta\Phi_1+\delta\Phi_2+\delta\Phi_3)^2/3$ in $L_{\sigma,\Phi}$ and $2\lambda(\delta\sigma)^2\vec{\pi}^2$ in $L_{\pi,\sigma}$, 
we can see there are two one loop amplitudes of the decay, 
which are proportional to $2\lambda h\sin^3\theta\cos^2\theta$ and $2\lambda h\sin\theta\cos^4\theta$, respectively. 
Therefore, we find from the naive order estimation that the tree amplitude $\propto 4\lambda\sigma_0\sin\theta$ of the decay is dominant because $h/\sigma_0\simeq 0.2$
and  $2\lambda\cos^4\theta<1$.
In this way the monopole decays into pions through the mixing term between $\sigma$ field and monopole fields $\Phi_i$ caused by $L_{\sigma,\Phi}$.

\section{SU(3) linear sigma model coupled with monopoles}
\label{su(3)}

It is straightforward to generalize the SU(2) linear sigma model to SU(3) linear sigma model.
The model is composed of SU(3) meson field $\Sigma$ and the monopoles $\Phi_i$ described above.
Instead of the Lagrangian $L_{\Sigma}$, SU(3) meson field is described by the following Lagrangian,

\begin{eqnarray}
L_{su(3)\Sigma}&=&\frac{1}{4}\rm{Tr}(\partial_{\mu}\Sigma^{\dagger}\partial^{\mu}\Sigma)-\frac{m^2}{4}\rm{Tr}(\Sigma^{\dagger}\Sigma)
-\frac{\lambda'_1}{16} (\rm{Tr}(\Sigma^{\dagger}\Sigma))^2-\frac{\lambda'_2}{16}Tr((\Sigma^{\dagger}\Sigma)^2) \nonumber \\
&+&\frac{c}{8}(\det(\Sigma^{\dagger})+\det(\Sigma))+\frac{1}{2}\rm Tr(H(\Sigma^{\dagger}+\Sigma)),
\end{eqnarray}
where $\Sigma$ is composed of scalar $\sigma_a$ and pseudscalar $\pi_a$ components; $\Sigma\equiv\sum_{a=0\sim 8}(\sigma_a+i\pi_a)\lambda_a$
with Gell-Mann metrices $\lambda_i$ ( $i=1\sim 8$ ) and $\lambda_0=1\times \sqrt{2/3}$.
The fifth term proportional to $c$ represents QCD chiral anomaly and the last term corresponds to current quark masses of u, d and s quarks.
Assuming flavor SU(2) symmetry, we may put $H=h_0\lambda_0+h_8\lambda_8$. The term resembles a term of a spin system with external magnetic field $H$
imposed on spins $\Sigma$. The model has the symmetry SU$_V(3)\times $U$_A(3)$ when $H=0$ and $c=0$. In particular  the term proportional to $c$ 
breaks U$_A(1)$ symmetry associated with U$_A(1)$ chiral anomaly in QCD.  The parameter $m^2$ is positive so that there are no vacuum condensation,
i.e. $\langle\Sigma\rangle=0$ as long as $h_{x,y}=0$. 

When the scalar fields $\sigma_0$ and $\sigma_8$ have nonzero vacuum expectation values,
they generate masses of quarks $q=(u,d,s)$, 

\begin{equation}
g'\bar{q}(\langle\sigma_0\rangle\lambda_0+\langle\sigma_8\rangle\lambda_8) q=m_{ud}(\bar{u}u+\bar{d}d)+m_s\bar{s}s
\end{equation}
with $m_{ud}\equiv g'(\langle\sigma_0\rangle\sqrt{2/3}+\langle\sigma_8\rangle\sqrt{1/3})= g'\langle\sigma_x\rangle$ 
and $m_s\equiv g'(\langle\sigma_0\rangle\sqrt{2/3}-\langle\sigma_8\rangle\sqrt{4/3})=\sqrt{2}g'\langle\sigma_y\rangle$;
$\sigma_x\equiv \sigma_0\sqrt{2/3}+\sigma_8\sqrt{1/3}$ and $\sigma_y\equiv \sigma_0\sqrt{1/3}-\sigma_8\sqrt{2/3}$.
The coupling between quarks $q$ and the scalar fields $\sigma$ is taken in standard analysis\cite{che} of the SU(3) linear sigma model. 
But we do not address the interaction in the present model. As we have discussed, the current quark masses are generated by the
monopole condensation $\langle\Phi\rangle$, not chiral condensate $\langle\sigma\rangle$.

\vspace{0.1cm}
The monopoles $\Phi_i$ couple with sigma field $\sigma$ in the SU(2) model. In the SU(3) model
they couple with isoscalar scalar mesons $\sigma_x$ and $\sigma_y$, which is described by the following Lagrangian,

\begin{equation}
\label{20}
L_{int}^{SU(3)}=h\sum_{i=1\sim 3}|\Phi_i|^2(\sigma_x+\sqrt{2}\sigma_y)
\end{equation}  
where we have taken into account the fact that the monopole quark interactions are independent of flavors, for instance,
$\sum_{i\sim 3}|\Phi_i|^2(\bar{u}u+\bar{d}d+\bar{s}s)$.
The monopoles $\Phi_i$ are described by the dual superconducting model $L_{\Phi}$ given above.
Therefore, the SU(3) linear sigma model coupled with the monopoles is described by the Lagrangian $L_{su(3) \Sigma}+L_{\Phi}+L_{int}^{SU(3)}$.

\vspace{0.1cm}
First, we would like to see vacuum configurations of the scalar fields $\sigma_x$, $\sigma_y$ and the monopole fields $\Phi_i$. Their potential is
given by

\begin{eqnarray}
V(\sigma,\Phi)&=&\frac{m^2}{2}(\sigma_x^2+\sigma_y^2)+\frac{\lambda_1'}{2}\sigma_x^2\sigma_y^2+\frac{2\lambda_1'+\lambda_2'}{8}\sigma_x^4
+\frac{\lambda_1'+\lambda_2'}{4}\sigma_y^4-\frac{c}{2\sqrt{2}}\sigma_x^2\sigma_y \nonumber \\
-h_x\sigma_x&-&h_y\sigma_y 
+ \sum_{i=1\sim 3}(-\mu^2|\Phi_i|^2+\lambda'|\Phi_i|^4)-h\sum_{i=1\sim 3}|\Phi_i|^2(\sigma_x+\sqrt{2}\sigma_y)
\end{eqnarray}
where the term with $h_x$ represents $u$ and $d$ quark bare mass so that we may approximately put $h_x=0$.
We only retain the term with $h_y$ which generates the $s$ quark mass.
The term with $c$ represents QCD chiral anomaly. Here we put $c=0$. The anomaly is already taken into account in the monopole meson interactions in our model.
As we have explained, quarks change their chiralities when they collide with the monopoles. 
The chirality non conservation is caused\cite{rubakov,callan,ezawa} by
the chiral anomaly. Thus, the monopole hadron interaction is produced by the chiral anomaly. For simplicity we assume that the anomaly is only present
in the interaction term $L_{int}^{SU(3)}$.

\vspace{0.1cm}
The vacuum configurations of these fields are determined by solving following equations

\begin{eqnarray}
\label{22}
0=\frac{\partial V}{\partial \sigma_x}&=&m^2\sigma_x+\lambda_1'(\sigma_x^2+\sigma_y^2)\sigma_x+\frac{\lambda_2'}{2}\sigma_x^3-h\sum_{i=1\sim 3}|\Phi_i|^2 \\
\label{23}
0=\frac{\partial V}{\partial \sigma_y}&=&m^2\sigma_y+\lambda_1'\sigma_x^2\sigma_y+(\lambda_1'+\lambda_2')\sigma_y^3-\sqrt{2}h\sum_{i=1\sim 3}|\Phi_i|^2-h_y  \\
0=\frac{\partial V}{\partial \Phi_i^{\dagger}}&=&(-\mu^2+2\lambda'|\Phi_i|^2)\Phi_i-h\Phi_i(\sigma_x+\sqrt{2}\sigma_y).
\end{eqnarray}
  
As in the SU(2) linear sigma model, without the explicit breaking term $\propto h_y$,
the chiral condensate ( $\langle\sigma_x\rangle\neq 0$ and $\langle\sigma_y\rangle\neq 0$ ) arises
only when the monopole condensation $\langle\Phi_i\rangle\neq 0$ takes place. 
From the above equations, we can see that when $h_y=0, \sigma_{x,y}\to 0$ as $\Phi_i\to 0$ 
because all parameters $m^2$, $\lambda_{1,2}'$ are positive.
Because we have the term with $h_y\neq 0$ leading to the $s$ quark mass,
this explicit chiral symmetry breaking term gives $\langle\sigma_y\rangle\neq 0$ without the monopole condensation.

By shifting the fields such as $\sigma_x=\bar{\sigma}_x+\delta\sigma_x$ and $\sigma_y=\bar{\sigma}_y+\delta\sigma_y$ with $\bar{\sigma}_x\equiv\langle\sigma_x\rangle$
and $\bar{\sigma}_y\equiv\langle\sigma_y\rangle$,
we can obtain masses of mesons like pions and kaons. These masses are identical to those derived from
the Lagrangian $L_{su(3)\Sigma}$. The mixing term in eq(\ref{20}) does not contribute to the masses. The term mixes the monopoles with
$\sigma_{x,y}$ fields representing $\sigma$ and $f_0$ mesons.
Interestingly, we can find that the mixing parameter $h$ is much small.
Because the mass of the pions is given by

\begin{equation}
m_{\pi}^2=m^2+\lambda_1'(\bar{\sigma}_x^2+\bar{\sigma}_y^2)+\lambda_2'\bar{\sigma}_x^2,
\end{equation}
we find from eq(\ref{22}) that 

\begin{equation}
m_{\pi}^2=\frac{3h|\Phi_0|^2}{\bar{\sigma}_x},
\end{equation}
with $\langle\Phi_i\rangle=\Phi_0$. This is just the one in eq(\ref{10}) derived from the SU(2) linear sigma model. Because $\bar{\sigma}_x$ represents 
the pion decay constant, $f_{\pi}\simeq 92$MeV, we see that the mixing parameter $h\simeq 20$MeV is small compared with typical hadronic scale $100$MeV.

\vspace{0.1cm}
Diagonalizing the mixing term, we can obtain the masses of the monopoles and scalar mesons $\sigma$ and $f_0$. But
the explicit formulas of the masses are very complicated in comparison with the ones in the SU(2) linear sigma model. 
Here we would like to show the masses in the limit of the vanishingly small mixing parameter $h$.

When $h=0$, we have $\bar{\sigma}_x=0$ and $\Phi_0=\sqrt{\mu^2/2\lambda'}$. Thus, the masses of the monopoles, $\sigma_x$ and $\sigma_y$ mesons 
are given by 

\begin{equation}
M_{mon}^2=4\mu^2, \quad M_{\sigma_x}^2=m^2+\lambda_1'\bar{\sigma}_y^2 \quad \mbox{and}  \quad M_{\sigma_y}^2=m^2+3(\lambda_1'+\lambda_2')\bar{\sigma}_y^2,
\end{equation}
respectively, where $\bar{\sigma}_y$ satisfies the formula $(m^2+(\lambda_1'+\lambda_2')\bar{\sigma}_y^2)\bar{\sigma}_y=h_y$.
In order to diagonalize the mixing term we extract the quadratic terms of the fields  $\delta\sigma_x$, $\delta\sigma_y$ and $\delta\Phi_i$ ( $\Phi_i=\Phi_0+\delta\Phi_i$ )
from the Lagrangian,

\begin{equation}
\label{28}
\frac{A}{2}(\delta\sigma_x)^2+\frac{B}{2}(\delta\sigma_y)^2+\frac{C}{2}\sum_{i=1\sim 3}(\delta\Phi_i)^2-\frac{D}{2}(\delta\sigma_x+\sqrt{2}\delta\sigma_y)\sum_{i=1\sim 3}\delta\Phi_i
+E\delta\sigma_x\delta\sigma_y
\end{equation} 
where the parameters $A$, $B$, $C$ $D$ and $E$ are given by

\begin{eqnarray}
A&\equiv&m^2+\lambda_1'(\bar{\sigma}_y^2+2\bar{\sigma}_y\delta\bar{\sigma}_y), \quad B\equiv m^2+3(\lambda_1'+\lambda_2')(\bar{\sigma}_y^2+2\bar{\sigma}_y\delta\bar{\sigma}_y),
 \nonumber \\ 
C&\equiv&4\mu^2+24\Phi_0\lambda'\delta\Phi_0-2\sqrt{2}h\bar{\sigma}_y \quad
D\equiv 4h\Phi_0, \quad E\equiv2 \lambda'\delta\bar{\sigma}_x\bar{\sigma}_y
\end{eqnarray}
with $\delta\bar{\sigma}_x=3h|\Phi_0|^2/(m^2+\lambda_1'\bar{\sigma}_y^2)$, $\delta\bar{\sigma}_y=3h\Phi_0^2/(m^2+3(\lambda_1'+\lambda_2')\bar{\sigma}_y^2)$ 
and $\delta\Phi_0=\sqrt{2}h\bar{\sigma}_y/4\lambda'\Phi_0$,
where we have taken the parameters up to the order of $h$.

Then, by diagonalizing eq(\ref{28}) we obtain the masses up to the order of $h$.
They are given by

\begin{eqnarray}
\label{30}
M_{mon}^2&=&4\mu^2+24\lambda'\Phi_0\delta\Phi_0-2h\sqrt{2}\bar{\sigma}_y=4\mu^2+4\sqrt{2}h\bar{\sigma}_y \\
\label{31}
M_{\bar{\sigma}_x}^2&=&m^2+\lambda_1'(\bar{\sigma}_y^2+2\bar{\sigma}_y\delta{\sigma}_y)=
m^2+\lambda_1'(\bar{\sigma}_y^2+\frac{6h\bar{\sigma}_y\Phi_0^2}{m^2+3(\lambda_1'+\lambda_2')\bar{\sigma}_y^2}) \\
\label{32}
M_{\bar{\sigma}_y}^2&=&m^2+3(\lambda_1'+\lambda_2')(\bar{\sigma}_y^2+2\bar{\sigma}_y\delta\bar{\sigma}_y)
=m^2+ \nonumber \\
&+&3(\lambda_1'+\lambda_2')(\bar{\sigma}_y^2+\frac{6h\bar{\sigma}_y\Phi_0^2}{m^2
+3(\lambda_1'+\lambda_2')\bar{\sigma}_y^2})\, .
\end{eqnarray}

These results are obtained in the limit of the vanishingly small mixing parameter $h$. For the parameter $h\simeq 20$MeV, we need numerical
calculations for obtaining the masses. But, we expect that the mass hierarchy $M_{\sigma_y}(h)>M_{\sigma_x}(h)$ holds even for the value $h\simeq 20$MeV.
Thus,
we may identify the masses such that
$M_{\sigma_x}$ ( $M_{\sigma_y}$ ) denotes the mass of the sigma meson ( $f_0$ meson ), while
$M_{mon}$ does the monopole.

\vspace{0.1cm}
Now we would like to discuss the decay of the monopole with the mass $M_{mon}$. The monopole is mixed with the scalar fields $\sigma_{x,y}$;
they involve the component of the monopole. That is, the fields $\delta\sigma_{x,y}$ involve the monopole with mass $M_{mon}$, whose fraction is given by

\begin{equation}
\label{33}
\frac{\delta\sigma_x}{\delta\sigma_y}=\frac{4\mu^2-m^2-3(\lambda'_1+\lambda'_2)\bar{\sigma}_y^2}{\sqrt{2}(4\mu^2-m^2-\lambda'_1\bar{\sigma}_y^2)}=
\frac{M_{mon}^2(h=0)-M_{\bar{\sigma}_y^2}(h=0)}{\sqrt{2}(M_{mon}^2(h=0)-M_{\bar{\sigma}_x^2}(h=0))}<1
\end{equation}
up to the order of $h$ where we have assumed that $M_{mon}>M_{\bar{\sigma}_y}>M_{\bar{\sigma}_x}$.
The inequality implies that the monopole couples with hadrons mainly through the mixing with the field $\delta\sigma_y$.
The scalar field couples with ordinary mesons, pion, kaon, $\eta$, $\eta'$ etc. through the Lagrangian $L_{su(3)\Sigma}$. 
But it couples more strongly with kaons than the field $\delta\sigma_x$ does.
Therefore, the decay modes
of the monopole are expected to have relatively large fraction of the $\bar{K}K$ decay in comparison to the decay modes of ordinary mesons
composed of $\bar{u}u$ and $\bar{d}d$. 
Furthermore, there is no
two gamma decay of the monopole. On the other hand, the mesons composed of $\bar{u}u$ $\bar{d}d$ or $\bar{s}s$ may have comparatively
large probability of the two gamma decay. Under these consideration, we may identify the monopole as $f_0(1500)$ meson.  

\section{Summary and conclusion}
\label{con}
Assuming that the monopoles are real dynamical variables in strong coupled QCD, we have examined monopole quark interaction.
We have shown that the quarks must change their chiralities when they collide with the monopoles. 
That is, the monopoles break the chiral symmetry SU$_A$(2)$\times$ U$_A$(1). The chirality non conservation arises at the locations of the monopoles
through the chiral non conserved boundary condition. The boundary condition is effectively represented by 
the interaction $-g'\sum_{i=1\sim 3}|\Phi_i|^2(\bar{u}u+\bar{d}d+\bar{s}s)$. It describes the quark scattering at the location of the monopole.
It also leads to current quark masses when the monopole condensation takes place, i.e. $\langle\Phi\rangle\neq 0$.
( Owing to this term, Gell-Mann-Oaks-Renner relation is modified such as $m_{\pi}^2f_{\pi}^2=-(2m_{ud}+3g'\Phi_0^2)\langle\bar{q}q\rangle$. )   
Therefore, pions are not Nambu-Goldstone bosons even in the vanishing bare quark masses $m_{ud}=0$. 
The small masses of the pions arise mainly from the monopole quark interaction. The interaction is weak relative to the quark gluon interactions.
This is because quarks have color charge $g$ and the monopoles have color magnetic charges $g_m=1/2g$. 

The effective monopole quark interaction gives rise to the chiral condensate\cite{iwazaki2} $\langle\bar{q}q\rangle<0 $ as well as the current quark masses $3g'\Phi_0^2/2$  
when the monopoles condense. That is,
the masses of hadrons are generated by the monopole condensation. The masses are expected to decrease in dense nuclear matter because
the monopole condensate is expelled by the stronger color electric fields in the matter than those in hadrons.  
The intimate relation between the monopole condensation and the chiral condensation is consistent with the previous analysis that
the local chiral condensate $\langle\bar{q}q\rangle\propto 1/r^3$ \cite{ezawa} is present around each monopole at $r=0$.

\vspace{0.1cm} 
Based on the analysis of the monopole quark interaction, 
we have proposed a phenomenological SU(3) ( SU(2) ) linear sigma model coupled with the QCD monopoles. 
The monopoles are described by dual superconducting model.
They couple only with isoscalar scalar mesons $\sigma_{x,y}$ ( $\sigma$ ). 
The coupling is given by $h\sum_{i=1\sim 3}|\Phi_i|^2(\sigma_x+\sqrt{2}\sigma_y)$ ( $h\sigma \sum_{i=1\sim 3}|\Phi_i|^2$ ).
It corresponds to the monopole quark interaction. It explicitly breaks SU$_{A}$(2) symmetry,
and mixes the monopoles and the mesons. 
Until now, we have had no guiding principle for the coupling between the monopoles and hadrons. But  
the analysis of the monopole quark interaction uniquely leads to the coupling. 
Thus, it is very intriguing to acquire the appropriate coupling or the mixing between the monopoles and hadrons.  

The linear sigma model in the present paper does not give rise to the spontaneous generation of chiral condensate,
when the monopole hadron coupling is absent.
The condensate arises through the coupling only when the monopole condensation takes place. In other words, 
the chiral condensate $\langle\sigma\rangle\neq 0$ vanishes when the monopole condensate $\langle\Phi\rangle\neq 0$ vanishes.
This generation mechanism of the chiral condensate comes from the basic analysis\cite{iwazaki2} that the monopole quark
interaction leads to the chiral condensate when the monopole condenses.


\vspace{0.1cm}
Because the chiral symmetry is explicitly broken in the coupling, the pions
are not massless.
We have derived the relation between pion mass and the monopole condensate $\langle\Phi_i\rangle=\Phi_0$, 
that is, $m_{\pi}^2=3h|\Phi_0|^2/f_{\pi}$.
We find from the relation that
that the coupling $h$ is relatively small $h\sim 20$MeV. 
The smallness of the coupling, in other words, the mixing between the monopole and hadrons is consistent with the result obtained in lattice gauge theories;
glueball mass with the lowest energy obtained in quenched approximation does not change so much even if dynamical fermions are taken into account. 

By diagonalizing the small mixing term,
we have obtained up to the order of $h$ the masses of $\sigma$ meson, $f_0$ meson and the observable color singlet monopole.
The monopole decays into pions, kaons and other hadrons through the mixing term. Namely, the sigma fields $\sigma_x$ and $\sigma_y$ 
involve the monopole component and couple with
the other hadrons in the way controlled by the sigma model. 
( The scalar field $\sigma_y$ involves the larger amount of the monopole component than the field $\sigma_x$ does. )
Thus, the monopole decays into the hadrons in the same way 
as isoscalar scalar mesons composed of $\bar{q}q$ decay.
But the monopole does not decay into $2\gamma$.
Therefore, it is natural to identify the monopole with $f_0(1500)$ meson because it decays into two pions, four pions, kaons and two $\eta$
but does not decay into $2\gamma$.
 
 In summary, we have presented phenomenological consequences based on our idea 
 that the chiral symmetry is explicitly broken by the monopoles in the strong coupled QCD, while it is not broken
 in the weak coupled QCD.

\section*{Acknowledgment}

 \vspace{0.2cm}
The author
expresses thanks to Prof. O. Morimatsu KEK, for useful comments
and discussions.



\end{document}